# Modeling degradation of Lithium-ion batteries for second-life applications: preliminary results

Gabriele Pozzato, *Member, IEEE*, Seong Beom Lee, and Simona Onori, *Senior Member, IEEE*

*Abstract*— **This paper presents a novel battery modeling framework based on the enhanced single particle model (ESPM) to account for degradation mechanisms of retired electric vehicle batteries. While accounting for the transport and electrochemical phenomena in the battery solid and electrolyte phases, the dominant anode-related aging mechanisms, namely, solid electrolyte interphase (SEI) layer growth and lithium plating, are modeled. For the first time, the loss of active material (LAM), which describes the tendency of anode and cathode, over time, to reduce the electrode material available for intercalation and deintercalation, is introduced in the ESPM. Moreover, the coupling of the aging mechanisms with the LAM dynamics provides a comprehensive means for the prediction of both linear and non-linear capacity fade trajectories, crucial to assess the health of batteries that are considered for second-life applications. Relying on data borrowed from [13], a model parameter identification and a comprehensive sensitivity analysis are performed to prove the effectiveness of the modeling approach.**

I. INTRODUCTION

Lithium-ion battery (LIB) technology takes advantage of the high electrochemical potential of lithium (-3.040V vs. standard hydrogen electrode) while providing high specific power (300-1500W/kg) and specific energy density (100-250Wh/kg) against all the electrochemical battery devices available on the market [1]. For this reason, LIBs are considered the best means to store energy to support sustainable transportation systems and renewable grid applications. In 2019, the market of LIBs was valued at $36.7 billion and is expected to grow by an annual rate of 18.0% to reach $129.3 billion by 2027 [2].

Deployment of electric vehicles (EVs) has been growing rapidly over the past decade. It was estimated that 1.18 million EVs were present on the U.S. roads in 2019 [4]. Typically, EV batteries have a lifetime of 8-10 years, and they retire from the first life when their capacity and/or power capability decreases to 70-80 percent of their initial values. The first large wave of retired EV batteries is expected by 2025, after the 15-year lifetime of the Tesla Model S and Nissan LEAF that first went to market [6]. In California, about 45,000 EV batteries are expected to retire by 2027 [7].

Potential end-of-life pathways for retired EV batteries include repurposing, remanufacturing, recycling, and disposal [8]. '*Repurposing*', or reuse, is defined as the redeployment of retired EV batteries in a different application from the one they were first used. Retired EV batteries are postulated to be able to provide energy storage services in a stationary grid application. In the '*Remanufacturing*' process, damaged battery cells/module are replaced with new ones within the battery module/pack and reused in EVs. The objective of '*Remanufacturing*' is to extend the useful life of batteries with minimal additional cost. In '*Recycling*' extraction and reprocessing of valuable metals (e.g., nickel and cobalt) are practiced. A sustainable life cycle would apply the '*Recycling*' process after the batteries have been repurposed to provide a second-life. In the '*Disposal*' pathway, the battery cells are moved to landfill and discarded under the assumption that they are no longer worth recycling.

Compared to other routes, the repurposing of EV batteries would maximize the total lifetime value and revenues of the device. The optimal deployment of repurposed retired EV batteries holds potential to reduce up-front costs and provides benefits to consumers and utilities, maximizing the usability and performance of the device.

To date, most second-life EV battery projects have been conducted by industrial vendors. In 2013, for example, ABB and General Motors built a 25kW energy storage system in San Francisco, California, USA by collecting retired EV batteries from the Chevrolet Volt plug in hybrid vehicle [10]. In 2015, BMW, Vattenfall, and Bosch obtained retired batteries from more than a hundred EVs and jointly constructed a 2 MW, 2800 kWh second-life battery energy storage system for grid support (Hamburg, Germany) [11]. In the same year, Toyota also built a stand-alone 10MW energy storage system with Prius retired batteries at the Lamar Buffalo Ranch in Yellowstone National Park, USA. This storage system supported a wind-battery microgrid system [12]. These early-stage research projects provided no systematic approach for the deployment of second-life batteries in grid applications; the original battery packs are just collected as they are and deployed in a microgrid or a building site to supply the needed power and energy.

A battery cell is a basic unit of a LIB whereas a battery module contains multiple battery cells, connected either in series or in parallel. Similarly, a battery pack involves multiple battery modules in parallel and/or series. In a battery module, individual cells packed closely together interact with each other thermally, and heat is transferred through conduction between the surface of neighboring cells [19].

G. Pozzato (e-mail: gpozzato@stanford.edu), S. B. Lee (e-mail: sblee84@stanford.edu), and S. Onori (corresponding author, phone: 650-498-1183; e-mail: sonori@stanford.edu) are with the Department of Energy Resources Engineering, Stanford University, Stanford, CA 94305 USA

The SOH estimation and cell screening approaches for second-life batteries found in the literature are based on the use of empirical models and lack electrochemical dynamics information. To date, the SEI layer growth has been regarded as the main cause of capacity fade and impedance increase of the battery system [13]. Accurate estimation of lithium plating dynamics is essential to assess future and safe usability of retired EV batteries. Lithium plating is defined as the formation of metallic lithium on the anode of LIBs during charging ($Li^{+}+e^{-} \rightarrow Li$). Lithium plating not only causes capacity and power fade but also poses significant safety concerns when it forms unmonitored. Lithium plating can lead to irregular dendrite-shaped growth that could cause thermal runaway inside the device. Once EV batteries are retired, one critical question is whether lithium plating has developed. Typically, lithium plating is known to occur under extreme operating charging conditions, such as fast charging and low temperature [9]. In order to study the second-life of retired EV batteries, it is deemed important to investigate whether lithium plating is generated under operating conditions controlled by the battery management system (BMS). Several nondestructive lithium plating detection methods have been proposed for online battery characterization.

In [14], differential voltage analysis to detect the high voltage plateau was introduced to provide quantitative results concerning the lithium stripping reaction. Lithium stripping, which is the opposite reaction ($Li \rightarrow Li^{+}+e^{-}$) to lithium plating, can occur during discharge after lithium plating has formed during charge. During discharge, if lithium plating already exists on the anode, the metallic lithium favors the conversion to lithium ions rather than deintercalation of lithium ions inside the solid particles (the electrochemical potential of metallic lithium is higher compared to the potential of lithium deintercalation from the negative electrode). For this reason, a voltage plateau can be observed at the initial stage of the discharge process, when starting from a fully charged battery. In [14], a cylindrical 26650-type commercial Li-ion battery ($LiFePO_4$) with 2.5 Ah capacity was tested under various low temperatures (-20°C, -22°C, -24°C, -26°C), within a SOC range of 100-10% and charge currents 1C C/2). This study showed that the differential voltage and capacity discharge curves can be used to calculate lithium plating quantitatively. In [13], the identification and quantification of lithium plating on a commercial graphite NMC cell was proposed. This study indicated that lithium plating can also occur at mild charging conditions after extended cycling, leading to rapid aging of the cell. This is a fundamental breakthrough in the field of SOH estimation. In addition, in [14-15] the loss of active material (LAM) was indicated as one of the causes leading to lithium plating after extended cycles (LAM is defined when active mass of the electrodes is no longer available for the insertion of lithium). The amount of lithium ions can exceed the accommodation of anode capacity due to the LAM of the electrode. Excess lithium ions can plate on the surface of the graphite. Mathematical battery models can be used to understand and optimize LIBs' performance. Among many other models, physics-based battery models deliver internal physics dynamics such as transport and kinetic phenomena. Moreover, physics-based battery models can incorporate transport phenomena, chemical/electrochemical kinetics, side reactions, and thermal/stress/mechanical effects. Physics-based models offer the SOC and SOH along with internal electrochemical information that can be used to maximize safety, usability, and lifetime of the battery.

In this paper, we formulate a physics-based modeling framework that integrates LAM, lithium plating, and SEI dynamics, using the enhanced single particle model (ESPM). This paper consists of four sections: ESPM, Parameter identification, Results, and Conclusion. In the ESPM section, governing equations including the SEI layer growth, lithium plating, and LAM are presented. In the Parameter Identification section, the particle swarm optimization (PSO) algorithm is used to identify the kinetic/transport parameters. The experimental discharge voltage profiles at constant C-rate obtained from [13] are used to identify the model parameters. In the Results section, we show the model performance over experimental data from [13].

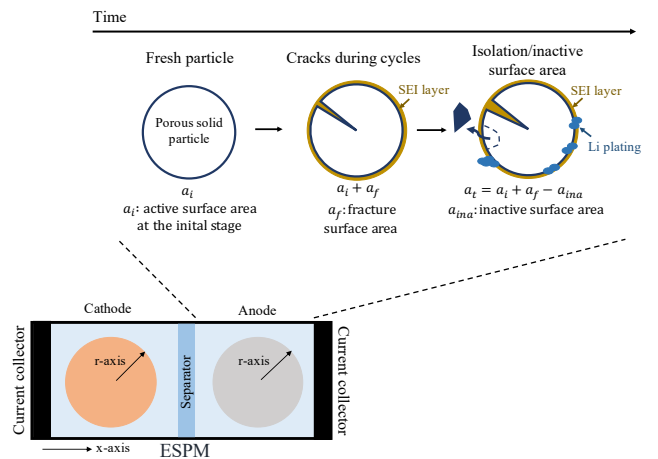

Figure 1. Schematic representation of the ESPM. Starting from a fresh particle, mechanical stresses due to lithium intercalation and deintercalation can induce the formation of cracks. After prolonged cycling, isolation of active material can also occur, together with SEI growth and lithium plating.

## II. ENHANCED SINGLE PARTICLE MODEL

The ESPM is formulated upon the assumption that each electrode can be represented by a single spherical particle, which implies that all solid particles are uniform and have the same chemical properties (see *Figure 1*) [17].

In the ESPM framework, solid particle concentration is described along with the radial dimension $r$. The variation of electrochemical potential in the solid particles along $x$ is ignored whereas the electrolyte concentration and potential are considered. According to [3], ESPM shows only a few millivolts error for discharge voltage profiles at constant C-rate $\leq 5C$ (compared to the Doyle-Fuller-Newman (DFN) model). For this reason, in this paper, the ESPM is selected as a modeling tool to include SEI layer growth and lithium plating (considered the dominant aging mechanisms [13]) and LAM.

Table I. Nomenclature (*p*: positive electrode, *s*: separator, *n*: negative electrode).

| Symbol | Variables | Units |
|---|---|---|
| $t_+$ | Transference number | - |
| $\alpha$ | Transfer coefficient | - |
| $brugg$ | Bruggeman coefficient | - |
| $\varepsilon_i, \varepsilon_{i,0}$ | Porosity and its initial condition ($i = p,n$) | - |
| $\nu_i, \nu_{i,filler}$ | Active volume fraction of solid phase and filler ($i = p,n$) | - |
| $\beta$ | Fraction of lithium plating converted into SEI | - |
| $\beta_i'$ | Inactive area evolution coefficient ($i = p,n$) | - |
| $k_i'$ | Fracture evolution coefficient ($i = p,n$) | - |
| $SOC_i$ | State of charge ($i = p,n$) | - |
| $\theta_i$ | Normalized lithium concentration ($i = p,n$) | - |
| $\theta_{i,0\%}, \theta_{i,100\%}$ | Reference stoichiometry ratio at 0% and 100% $SOC$ ($i = p,n$) | - |
| $t, x, r$ | Time and cartesian/radial coordinates | s,m |
| $R_i$ | Particle radius ($i = p,n$) | m |
| $L_i$ | Region thickness ($i = p,s,n$) | m |
| $L_{film}$ | Thickness of the surface film | m |
| $L_{Li}, L_{SEI}$ | SEI and lithium plating thickness | m |
| $A_{cell}$ | Cell cross sectional area | m$^2$ |
| $a_i$ | Specific surface area ($i = p,n$) | m$^2$/m$^3$ |
| $a_{i,f}$ | Specific fracture surface area ($i = p,n$) | m$^2$/m$^3$ |
| $a_{i,ina}$ | Specific inactive surface area ($i = p,n$) | m$^2$/m$^3$ |
| $a_{i,t}$ | Total specific surface area ($i = p,n$) | m$^2$/m$^3$ |
| $k_f$ | SEI side reaction kinetic constant | m/s |
| $D_i$ | Electrolyte phase diffusion coefficient ($i = p,s,n$) | m$^2$/s |
| $D_{s,i}$ | Solid phase diffusion coefficient ($i = p,n$) | m$^2$/s |
| $c$ | Electrolyte concentration | mol/m$^3$ |
| $c_{s,i}$ | Solid phase concentration ($i = p,n$) | mol/m$^3$ |
| $c_{Li}, c_{SEI}$ | Plated lithium and SEI concentration | mol/m$^3$ |
| $k_i$ | Reaction rate ($i = p,n$) | m$^{2.5}$/(mol$^{0.5}\cdot$s) |
| $J_i$ | Pore wall flux ($i = p,n$) | mol/(m$^3\cdot$s) |
| $I$ | Applied current | A |
| $i_{0,i}$ | Exchange current ($i = p,n$) | A/m$^2$ |
| $i_{0,lpl}$ | Lithium deposition exchange current | A/m$^2$ |
| $j_{int}, j_{SEI}, j_{lpl}$ | Intercalation and side current densities (SEI, lithium plating) | A/m$^3$ |
| $\kappa_i$ | Liquid phase conductivity ($i = p,s,n$) | S/m |
| $\kappa_{SEI}$ | SEI layer ionic conductivity | S/m |
| $R_{film}$ | Film resistance | Ω |
| $R_l$ | Lumped contact resistance | Ω |
| $R_{el}$ | Electrolyte resistance | Ω |
| $\Phi_{s,n}, \Phi_{e,n}$ | Solid and liquid phase potential at the anode | V |
| $\Delta\Phi_e$ | Diffusion overpotential | V |
| $\phi_e$ | Liquid phase potential | V |
| $U_i$ | Open circuit potential ($i = p,n$) | V |
| $V_{cell}$ | Cell voltage | V |
| $\eta_i$ | Overpotential ($i = p,n$) | V |
| $T, T_{ref}$ | Battery cell temperature and reference | K |
| $R$ | Universal gas constant | J/(mol$\cdot$K) |
| $E_{a,D_s}^i$ | Activation energy of solid diffusivity ($i = p,n$) | J/mol |
| $F$ | Faraday constant | C/mol |
| $M_{Li}, M_{SEI}$ | Plated lithium and SEI molar mass | kg/mol |
| $\rho_{Li}, \rho_{SEI}$ | Plated lithium and SEI density | kg/m$^3$ |

## A. Model governing equations

The ESPM model is in the form of partial differential equations (PDEs), including multivariable functions and their partial derivatives. The model predicts the dynamics of lithium concentration in the solid phase $c_{s,i}(x,r,t)$ at the anode and cathode (*i=p, n*), electrolyte concentration $c(x,t)$, and electrolyte potential $\phi_e(x,t)$. All equations, variables, and parameters are described in *Table I*, *II*, and *III*. Variables

Table II. Governing equations for the ESPM model. Only the boundary conditions modified with the lithium plating current density $j_{lpl}$ are shown. For a comprehensive description of the boundary conditions, the reader is referred to [17].

$$J_n = \frac{I}{A_{cell}FL_n}; \quad J_p = -\frac{I}{A_{cell}FL_p}; \quad J_s = 0$$

*Mass transport in the electrolyte phase* ($i = p, s, n$)
$$\varepsilon_i \frac{\partial c}{\partial t} = \frac{\partial}{\partial x}\left(D_{eff,i}(c,T)\frac{\partial c}{\partial x}\right) + (1-t_+)J_i$$

*Charge transport in the electrolyte phase* ($i = p, s, n$)
$$\kappa_{eff,i}(c,T)\frac{\partial}{\partial x}\left(\frac{\partial\phi_e}{\partial x}\right) + \frac{2\kappa_{eff,i}(c,T)RT}{F}(1-t_+)\frac{\partial^2 \ln c}{\partial x^2} + FJ_i = 0$$

*Mass transport in the solid phase* ($i = p, n$)
$$\frac{\partial c_{s,i}}{\partial t} = \frac{1}{r^2}\frac{\partial}{\partial r}\left(r^2 D_{s,i}(T)\frac{\partial c_{s,i}}{\partial r}\right); \quad \frac{\partial c_{s,i}}{\partial r}\bigg|_{r=0} = 0$$
$$\frac{\partial c_{s,n}}{\partial r}\bigg|_{r=R_n} = \frac{-I + L_n A_{cell}(j_{SEI} + j_{lpl})}{D_{s,n}(T)a_{n,t}A_{cell}FL_n}; \quad \frac{\partial c_{s,p}}{\partial r}\bigg|_{r=R_p} = \frac{I}{D_{s,p}(T)a_{p,t}A_{cell}FL_p}$$

and parameters are shown in *Table I*. Equations in *Table II* describe the battery charge and mass transport dynamics in the positive electrode, separator, and negative electrode. Additional equations are given in *Table III*. Equations T.3 show the porosity change due to aging phenomena. Equations T.4 describe the cell voltage as a function of the open-circuit voltage, overpotential, and liquid phase potential difference between the positive and negative electrodes, respectively. Ohmic losses due to the internal resistance and SEI layer growth are also included. Equations T.5 describe the electrochemical overpotential, and Equations T.6 and T.7 adopt the Butler-Volmer (BV) kinetic expression to model the intercalation/deintercalation of lithium ions, SEI layer growth, and lithium plating. The BV expression describes the charge transfer processes, which occur at the surface of solid particles–electrolyte interfaces (for SEI layer growth and lithium plating). The SEI layer growth dynamics was successfully implemented in our previous study [17], within the ESPM framework, following the approach proposed by [18]. In 2019, Narayanrao *et al.* modeled LAM by decreasing the specific surface area of solid particles from active surface area at the initial stage ($a_i$) [16]. This LAM model is established in two steps (see *Figure 1*); solid particles fracture ($a_{i,f}$) occurs first, resulting in a larger overall active surface area. The wider surface area at this stage is still active material, but as the surface area increases, the SEI layer is formed over a wider area, and the capacity loss phenomenon becomes more pronounced. The first step assumes that the ratio between the area created through the fracture to the total initial area of the electrode particles is a linearly increasing function of time as show below:

$$\frac{d\left(\frac{a_{i,f}}{a_i}\right)}{dt} = k_i' \rightarrow a_{i,f} = a_i k_i' \quad (1)$$

where $k_i'$ is the fracture rate coefficient.

In the second step, inactive surface area or isolation occurs under the assumption that the rate of increase of inactive area is proportional to the total surface area $a_{i,t}$:

$$\frac{da_{i,ina}}{dt} = \beta_i' a_{i,t} = \beta_i'(a_i + a_{i,f} - a_{i,ina}) \quad (2)$$

where $a_{i,ina}$ is the inactive surface area and $\beta_i'$ is the inactive area evolution coefficient. The inactive surface area dynamics (Equations T.8 and T.9 in *Table III*) can be obtained by

Table III. Additional equations for the ESPM model. The abbreviations *avg*, *eff*, *ref*, and *surf* stand for average, effective, reference, and surface.

*Diffusion and conduction coefficients (T.1)*
$D_{eff,i}(c,T) = D_i(c,T) \cdot \varepsilon_i^{brugg}$; $\kappa_{eff,i}(c,T) = \kappa(c,T) \cdot \varepsilon_i^{brugg}$; $i = p, s, n$
$D_{s,i}(T) = D_{s,i}^{ref} \exp\left[-\frac{E_{a,D_s}^i}{R}\left(\frac{1}{T} + \frac{1}{T_{ref}}\right)\right]$; $i = p, n$

*Active area (T.2)*
$a_i = 3/R_i$; $i = p, n$

*Porosity (T.3)*
$\varepsilon_{i,0} = 1 - \nu_i - \nu_{i,filler}$; $i = p, n$
$\varepsilon_p = \varepsilon_{p,0} + \frac{a_{p,ina} - a_{p,f}}{3} R_p$; $\varepsilon_n = \varepsilon_{n,0} + \frac{a_{n,ina} - a_{n,f}}{3} R_n - \nu_n \frac{3L_{film}}{R_n}$

*Cell voltage (T.4)*
$V_{cell} = U_p(\theta_p) - U_n(\theta_n) + \eta_p - \eta_n + \Delta\Phi_e - I(R_l + R_{el} + R_{film})$
$R_{el} = \frac{1}{2A_{cell}}\left(\frac{L_n}{\kappa_{eff,n}(c,T)} + \frac{2L_s}{\kappa_{eff,s}(c,T)} + \frac{L_p}{\kappa_{eff,p}(c,T)}\right)$
$\Phi_{s,n} = U_n(\theta_n) + \eta_n + R_{film}I$

*Electrochemical overpotential (T.5)*
$\eta_i = \frac{RT}{0.5F}\sinh^{-1}\left(\frac{I}{2A_{cell}a_{i,t}L_i i_{0,i}}\right)$; $i_{0,i} = k_i(T)F\sqrt{c^{avg}c_{s,i}^{surf}(c_{s,i}^{max} - c_{s,i}^{surf})}$
$i = p, n$

*Total mole flux (T.6)*
$\frac{I}{A_{cell}L_n} = j_{int} + j_{SEI} + j_{lpl}$

*Lithium plating and SEI layer growth (T.7)*
$j_{lpl} = -2a_{n,t}i_{0,lpl}\exp\left(-\frac{\alpha F}{RT}(\Phi_{s,n} - \Phi_{e,n} - R_{film}I)\right)$;
$j_{SEI} = -Fa_{n,t}k_f(c_{s,n},T)c_{solv}^{surf}\exp\left(-\frac{\alpha F}{RT}(\Phi_{s,n} - \Phi_{e,n} - R_{film}I)\right)$
$\frac{dc_{SEI}}{dt} = -\left(\frac{j_{SEI}}{2F} + \frac{j_{lpl}}{2F}\beta\right)$; $\frac{dc_{Li}}{dt} = -\frac{j_{lpl}}{2F}(1-\beta)$
$\frac{dL_{film}}{dt} = \frac{1}{a_{n,t}}\left(\frac{dc_{SEI}}{dt}\frac{M_{SEI}}{\rho_{SEI}} + \frac{dc_{Li}}{dt}\frac{M_{Li}}{\rho_{Li}}\right) = L_{SEI} + L_{Li}$; $R_{film} = \frac{L_{SEI}}{a_{n,t}A_{cell}L_n\kappa_{SEI}}$

*Loss of active material (T.8)*
$\frac{da_{i,ina}}{dt} = \beta_i'(a_i + a_{i,f} - a_{i,ina})$; $i = p, n$

*Total active area (T.9)*
$a_{i,t} = a_i + a_{i,f} - a_{i,ina}$; $a_{i,f} = a_i k_i' t$; $i = p, n$

*State of charge (T.10)*
$SOC_n = \frac{\theta_n - \theta_{n,0\%}}{\theta_{n,100\%} - \theta_{n,0\%}}$; $SOC_p = \frac{\theta_{p,0\%} - \theta_p}{\theta_{p,0\%} - \theta_{p,100\%}}$; $\theta_i = c_{s,i}^{surf}/c_{s,i}^{max}$; $i = p, n$

*Notes*
The variables $T$ and $\Delta\Phi_e$ are modeled as in [5]. The parameters $D_i(c,T)$, $\kappa(c,T)$, and $k_i(T)$ are computed according to [17]. Eventually, a comprehensive description of $k_f(c_{s,n},T)$ and $c_{solv}^{surf}$ is provided in [18].

combining Equations (1) and (2). This paper couples the inactive surface dynamics with related terms presented in *Table III*. Porosity (Equations T.3), electrochemical overpotential (Equations T. 5), and mole fluxes for lithium intercalation/deintercalation, SEI layer growth, and lithium plating are a function of the active surface area. As the total active surface area changes with cycles, the aforementioned electrochemical states vary. The total mole flux (Equation T.6) is a function of the mole fluxes for lithium intercalation/deintercalation, SEI layer growth, and lithium plating. As the total active area is reduced, the relative ratio of mole fluxes for lithium intercalation/deintercalation and SEI layer growth are decreased, while the relative ratio of mole flux for the lithium plating is increased.

## III. PARAMETER IDENTIFICATION

The identification of the model parameters is carried out relying on the particle swarm optimization (PSO) algorithm. To this aim, voltage *vs* capacity data for a 12.4Ah pouch cell discharged at C/3 (at 25°C) from [13] are used. In this study, the identification process is divided into two phases. First, the parameter vector $\Theta_1$ is identified over a fresh cell. Secondly, for the aged cell, the parameter vector $\Theta_2$ is identified. For the aged cell, experimental data at the 1000th and 3300th cycle are employed. Data at the 1000th cycle are related to aging conditions dominated by SEI growth. Conversely, measurements at the 3300th cycle are characterized by the superposition of two aging modes, namely, SEI growth and lithium plating. In accordance with [13], for both the fresh and aged cell, we assume that LAM is not occurring.

The identification aims at minimizing a cost function defined as follows:

$$J(\Theta) = w_1\sqrt{\frac{1}{N}\sum_{k=1}^{N}\left(V_{cell}(k;\Theta) - V_{cell}^{exp}(k)\right)^2} + \quad (3)$$

$$+ w_2\sqrt{\frac{1}{N}\sum_{k=1}^{N}\left(SOC_n(k;\Theta) - SOC^{exp}(k)\right)^2} + w_3\sqrt{\frac{1}{N}\sum_{k=1}^{N}\left(SOC_p(k;\Theta) - SOC^{exp}(k)\right)^2}$$

where $\Theta$ is a generic vector collecting the parameters to be identified ($\Theta_1$ and $\Theta_2$ for the fresh and aged cell, respectively), $SOC_p$, $SOC_n$ are the simulated state of charge at the cathode and anode (computed as in Equations T.10), $V_{cell}$ is the simulated voltage profile (Equations T.4), $k$ is the time instant, $N$ the number of samples, $V_{cell}^{exp}$ and $SOC^{exp}$ are the experimental cell voltage and state of charge from Coulomb counting, respectively. The weights $w_1$ [1/V], $w_2$ [-], and $w_3$ [-] are user-defined parameters here equal to one.

Starting from the fresh cell, the set of parameters to be identified is determined from the correlation analysis proposed in [17]. The values for the unidentified parameters are retrieved from [13] and [18] and are maintained unvaried moving from the fresh to the aged condition. The vector $\Theta_1$ contains the following parameters: the electrode area ($A_{cell}$), the lumped contact resistance ($R_l$), the volume fraction at the negative electrode ($\nu_n$), the solid particle radius at the positive and negative electrode ($R_p$ and $R_n$), the reference diffusion coefficients at the positive and negative electrode ($D_{s,p}^{ref}$ and $D_{s,n}^{ref}$), and the stoichiometry values at the positive and negative electrode ($\theta_{p,100\%}$ and $\theta_{n,100\%}$). Under this scenario, the SEI growth and lithium plating side reaction paths are inactive (no aging is happening on a fresh cell) and the identification is performed minimizing the objective function (3), with lower and upper bounds of the parameters defined in *Table IV*. The bounds are computed as ±50% of the initial guesses, with some exceptions: for the diffusion coefficients a larger range is analyzed, for $\nu_n$ ±20% is employed (values for the solid phase volume fraction lie around 0.5). Initial guesses for the identified parameters are obtained from [17] and [18]. For the aged cell, the parameter vector $\Theta_2$ is defined as the array of the following quantities: the ratio between the SEI thickness and ion conductivity ($L_{SEI}/\kappa_{SEI}$), and the stoichiometry values $\theta_{p,0\%}, \theta_{n,100\%}$. $L_{SEI}/\kappa_{SEI}$ is identified under the assumption of constant SEI layer during the discharge experiment, which is reasonable since $L_{SEI}$ is a slow varying variable. Generally speaking, open circuit potentials at the anode and cathode are a function of the normalized lithium concentration $\theta_i$, defined in Equations T.10. As SEI layer and lithium plating start growing, lithium ions participating in these side reactions are lost, leading to a reduction of the cyclable lithium and, consequently, to a modification of the solid phase normalized concentrations.

Therefore, as the battery cycles, the stoichiometry values $\theta_{i,0\%}$ and $\theta_{i,100\%}$ are expected to change. Among the four stoichiometry values, $\theta_{p,0\%}$ and $\theta_{n,100\%}$ are expected to vary more because of the SEI layer growth and lithium plating. In fact, if cyclable lithium is lost, during the discharge of the cell the anode open circuit potential increases more rapidly (with respect to the fresh condition) and $\theta_{n,0\%}$ is reached quickly, limiting the intercalation in the cathode. This leads to a reduction of the cathode solid phase lithium concentration (with respect to the fresh condition) or, equivalently, to a decreased $\theta_{p,0\%}$. Conversely, while charging, the cathode open circuit potential increases more rapidly (with respect to the fresh condition) and $\theta_{p,100\%}$ is quickly reached, limiting the intercalation in the anode. This leads to a decrease of the anode solid phase lithium concentration (with respect to the fresh condition) or, equivalently, to a decreased $\theta_{n,100\%}$. Therefore, $\theta_{p,0\%}$ and $\theta_{n,100\%}$ are the chosen parameters to be identified. Further details regarding this choice can be found in [15].

From experimental data at the 1000th cycle, the vector $\Theta_2$ is identified minimizing the cost function in Equation (3), with suitable upper and lower bounds for the parameters (*Table V*) and assuming the lithium plating side reaction to be inactive. A meaningful initial guess for $L_{SEI}/\kappa_{SEI}$ is obtained from [18]. Moreover, except for $\theta_{n,100\%}$, parameters in *Table IV* are not identified again. The outcome of the identification is summarized in the first column of *Table V*.

Table IV. $\Theta_1$ identified over a fresh cell.

| Parameter | Lower bound | Upper bound | Initial guess | Identified value | Unit |
|---|---|---|---|---|---|
| $A_{cell}$ | 0.28 | 0.83 | 0.55 | 0.57 | m² |
| $R_l$ | 0.02 | 0.07 | 0.045 | 0.04 | Ω |
| $\nu_n$ | 0.45 | 0.65 | 0.5 | 0.54 | - |
| $R_p$ | 0.6×10⁻⁶ | 1.9×10⁻⁶ | 1.25×10⁻⁶ | 10⁻⁶ | m |
| $R_n$ | 5×10⁻⁶ | 15×10⁻⁶ | 10×10⁻⁶ | 5.16×10⁻⁶ | m |
| $D_{s,p}^{ref}$ | 10⁻¹⁴ | 3.4×10⁻¹³ | 2.25×10⁻¹³ | 2×10⁻¹³ | m²/s |
| $D_{s,n}^{ref}$ | 10⁻¹⁴ | 3.4×10⁻¹³ | 2.25×10⁻¹³ | 10⁻¹³ | m²/s |
| $\theta_{p,100\%}$ | 0.14 | 0.41 | 0.28 | 0.30 | - |
| $\theta_{n,100\%}$ | 0.43 | 1 | 0.85 | 0.99 | - |
| **Cost function value: $J(\Theta_1) = 0.03$** | | | | | |

Table V. $\Theta_2$ identified over a cell aged for 1000 cycles.

| Parameter | Lower bound | Upper bound | Initial guess | Identified value | Unit |
|---|---|---|---|---|---|
| $L_{SEI}/\kappa_{SEI}$ | 0.0015 | 0.15 | 0.076 | 0.085 | Ωm² |
| $\theta_{p,0\%}$ | 0.7 | 1 | 0.85 | 0.92 | - |
| $\theta_{n,100\%}$ | 0.7 | 1 | 0.85 | 0.88 | - |
| **Cost function value: $J(\Theta_2) = 0.03$** | | | | | |

Table VI. $\Theta_2$ identified over a cell aged for 3300 cycles.

| Parameter | Lower bound | Upper bound | Initial guess | Identified value | Unit |
|---|---|---|---|---|---|
| $L_{SEI}/\kappa_{SEI}$ | 0.003 | 0.3 | 2 | 0.25 | Ωm² |
| $\theta_{p,0\%}$ | 0.6 | 1 | 0.8 | 0.79 | - |
| $\theta_{n,100\%}$ | 0.6 | 1 | 0.8 | 0.72 | - |
| **Cost function value: $J(\Theta_2) = 0.04$** | | | | | |

To reflect the increase in battery aging, $L_{SEI}/\kappa_{SEI}$, $\theta_{p,0\%}$, and $\theta_{n,100\%}$ must be updated with battery cycles. Therefore, the parameter vector $\Theta_2$ is identified again relying on experimental data at the 3300th cycle. At this stage of the life, the lithium plating reaction path is active, with the parameter $i_{0,lpl}$ retrieved from [13]. The identified parameters for the experimental data at the 3300th cycle are shown *Table VI*.

## IV. RESULTS

In *Figure 2*, simulated voltage and $SOC$ profiles are compared with experimental data at the 0th (fresh cell), 1000th, and 3300th cycle, respectively. Simulation results are obtained considering the corresponding identified parameters collected in *Tables IV*, *V*, and *VI*.

From Equation (2), the behavior of LAM is classified into three cases. Case (i): $k_i'$ and $\beta_i'$ are greater than zero and comparable in magnitude, i.e., $(1 - k_i'/\beta_i')$ is close to zero. Case (ii): $k_i'$ dominates over the inactive area evolution, i.e., $(1 - k_i'/\beta_i') \ll 0$. Case (iii): the inactive area evolution coefficient $\beta_i'$ dominates and $0 < (1 - k_i'/\beta_i') < 1$. Starting from values proposed in [16] and summarized in *Table VII* one can deduce that the cathode is mostly experiencing mechanical fractures, which increase the surface area available for intercalation/deintercalation reactions. Instead, the anode is experiencing LAM, which reduces the area available for both intercalation/deintercalation phenomena and side reactions.

Assuming the cathode active surface area to remain constant over time, i.e., neglecting the mechanical fractures, the impact of LAM dynamics on the cell behavior is analyzed. Thus, for each combination of parameters $k_n'$ and $\beta_n'$ in *Table VII*, the model is simulated and results are compared to experimental data in *Figure 3(a)*.

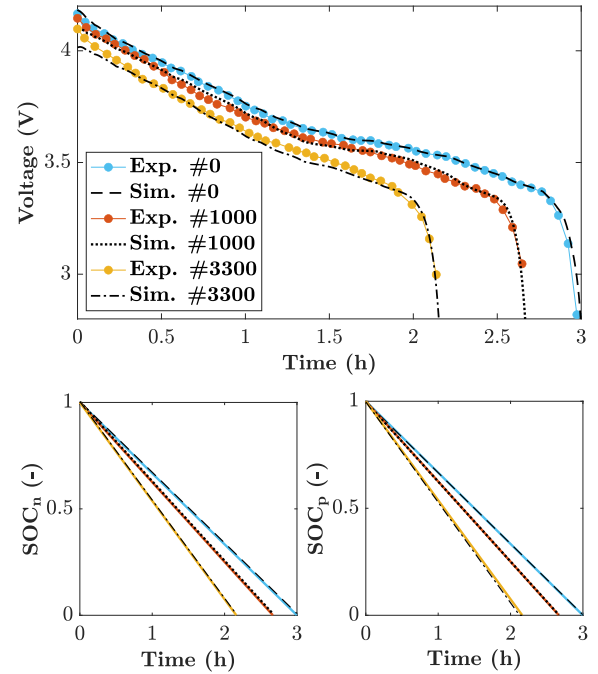

Figure 2. Comparison between experimental voltage (top) and $SOC$ profiles (bottom) with ESPM predicted voltage and $SOC$.

Table VII. Fracture and inactive area evolution coefficients. Minimum and maximum values are retrieved from [16] and remapped from cycle to time domain relying on the following expressions: $\beta'_i = \beta_i^{cycle}/T_{cycle}$, $k'_i = k_i^{cycle}/T_{cycle}$, with $T_{cycle}$ the time to perform a charge/discharge cycle.

| Cathode | $k_p^{cycle}$ | $k'_p$ | $\beta_p^{cycle}$ | $\beta'_p$ |
|---|---|---|---|---|
| | 1.65e-7 | 3.06e-11 | 1.07e-8 | 0.198e-11 |
| | 5e-7 | 9.26e-11 | 1e-7 | 1.85e-11 |
| Anode | $k_n^{cycle}$ | $k'_n$ | $\beta_n^{cycle}$ | $\beta'_n$ |
| | 7.57e-7 | 1.40e-10 | 4e-6 | 0.741e-9 |
| | 3.40e-6 | 6.30e-10 | 5.18e-5 | 9.59e-9 |

The reduction of the active area, related to an increment of the parameter $\beta'_n$, leads to a reduced discharged capacity. This phenomenon is mainly caused by the increment in the film resistance (Equations T.7 and *Figure 3(b)*), caused by the reduced total active surface $a_{n,t}$, and by the modification of the solid phase porosity (Equations T.3). Ultimately, the decreased discharged capacity is a symptom of the reduced capability of the anode of accepting lithium ions during the intercalation process. Thus, the negative electrode limits the charging capability of the cell, reducing the available capacity during the discharge process.

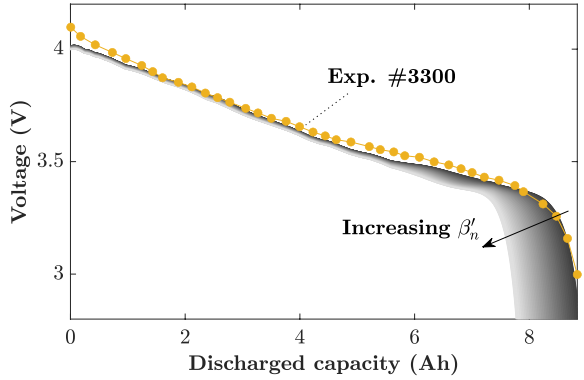

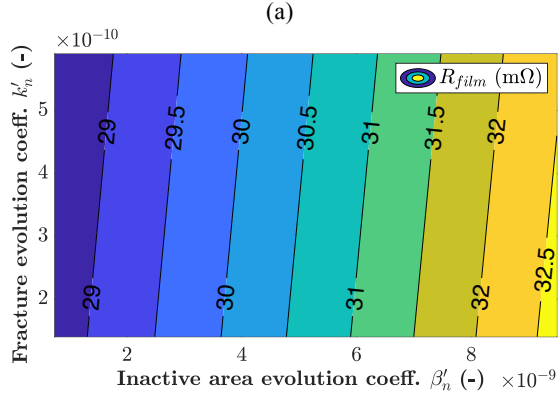

Figure 3. (a) Comparison of experimental voltage profile (yellow dots) with ESPM predicted voltages (grey shaded area) at the 3300th cycle as $\beta'_n$ varies within the limits reported in *Table VII*. (b) The film resistance for different combinations of the parameters $k'_n$ and $\beta'_n$ at the 3300th cycle.

## V. CONCLUSION

In this paper, the ESPM was formulated to include SEI growth, lithium plating, and LAM dynamics. LAM dynamics accounts for mechanical fractures, which lead to an increment of the active surface, and isolation, which brings to a decrement of the active surface. Coupling LAM dynamics with lithium plating can help to accurately predict the electrochemical state of retired EV batteries. Model identification results show the predictability of the proposed modeling strategy for 1) fresh cell scenario, aged conditions at 2) 1000th and 3) 3300th cycle). This model, a first to the authors' knowledge, will offer the opportunity to tackle the SOH evaluation and monitoring in retired batteries.

ACKNOWLEDGMENT

The research presented within this paper is supported by the Bits and Watts Initiative within the Precourt Institute for Energy at Stanford University.